\documentclass[10pt,conference]{IEEEtran}
\usepackage{cite}
\usepackage[xindy,style=index,numberedsection=false]{glossaries}
\usepackage{xurl}
\usepackage{graphicx}
\usepackage{xcolor}
\usepackage{bbold}
\usepackage{amsmath, amssymb}
\usepackage{cleveref}
\usepackage{subcaption}

\usepackage{booktabs}
\usepackage{makecell}
\usepackage{multirow}
\usepackage[binary-units=true]{siunitx} 

\usepackage{tikz}
\usepackage{circledsteps}

\usepackage[]{algpseudocode,algorithm}
\algtext*{EndWhile}
\algtext*{EndIf}
\algtext*{EndFor}
\algtext*{EndFunction}
\usepackage{setspace}
\let\Algorithm\algorithm
\renewcommand\algorithm[1][]{\Algorithm[#1]\setstretch{1.1}}
\algnewcommand{\LineComment}[1]{\State \(\triangleright\) #1}

\usepackage{booktabs}
\usepackage{makecell}
\usepackage{caption}
\newcommand{\ra}[1]{\renewcommand{\arraystretch}{#1}}

\newacronym{sae}{SAE}{Sparse Approximate Eigenproblem}
\newacronym{evd}{EVD}{Eigenvalue Decomposition}
\newacronym{ff}{FF}{Flip Flop}
\newacronym{lut}{LUT}{Look Up Tables}
\newacronym{sa}{SA}{Systolic Array}
\newacronym{nlp}{NLP}{Natural Language Processing}
\newacronym{ppr}{PPR}{Personalized PageRank}
\newacronym{spmv}{SpMV}{Sparse Matrix-Vector Multiplication}
\newacronym{topkspmv}{Top-K SpMV}{Top-K Sparse matrix-vector multiplication}
\newacronym{coo}{COO}{Coordinate}
\newacronym{dsl}{DSL}{Domain-Specific Language}
\newacronym{slr}{SLR}{Super Logic Region}
\newacronym{csc}{CSC}{Compressed Sparse Column}
\newacronym{csr}{CSR}{Compressed Sparse Row}
\newacronym{raw}{RAW}{Read-After-Write}
\newacronym{ndcg}{NDCG}{Normalized Discounted Cumulative Gain}
\newacronym{ir}{IR}{Information Retrieval}
\newacronym{dcg}{DCG}{Discounted Cumulative Gain}
\newacronym{er}{ER}{Entity Resolution}
\newacronym{ii}{II}{Initiation Interval}
\newacronym{gpu}{GPU}{Graphics Processing Unit}
\newacronym{fpga}{FPGA}{Field Programmable Gate Array}
\newacronym{colamd}{COLAMD}{ColumnApproximate Minimum Degree algorithm}
\newacronym{cpu}{CPU}{Central Processing Unit}
\newacronym{hbm}{HBM}{High Bandwidth Memory}
\newacronym{ddr}{DDR}{Double Data Rate}
\newacronym{bscsr}{BS-CSR}{Block-Streaming CSR}
\newacronym{glove}{GloVe}{Global Vectors for Word Representation}
\newacronym{cu}{CU}{Compute Unit}
\newacronym{pe}{PE}{Processing Element}
\newacronym{sp}{SP}{Stream Processor}
\newacronym{uram}{URAM}{UltraRAM}
\newacronym{cordic}{CORDIC}{Coordinate Rotation Digital Computer}
\newacronym{iram}{IRAM}{Implicitly Restarted Arnoldi Method}

\begin{document}

\title{Solving Large Top-K Graph Eigenproblems with a Memory and Compute-optimized FPGA Design}

\author{
\IEEEauthorblockN{Francesco Sgherzi\IEEEauthorrefmark{1}, Alberto Parravicini\IEEEauthorrefmark{2}, Marco Siracusa\IEEEauthorrefmark{1} Marco D. Santambrogio\IEEEauthorrefmark{2}}
\IEEEauthorblockA{Politecnico di Milano, DEIB, Milan, Italy}
\IEEEauthorrefmark{1}{\{francesco1.sgherzi, marco.siracusa\}}@mail.polimi.it
\IEEEauthorrefmark{2}{\{alberto.parravicini, marco.santambrogio\}}@polimi.it,
}

\newcommand{\XXX}{XXX}
\maketitle

\begin{abstract}
Large-scale eigenvalue computations on sparse matrices are a key component of graph analytics techniques based on spectral methods. 
In such applications, an exhaustive computation of all eigenvalues and eigenvectors is impractical and unnecessary, as spectral methods can retrieve the relevant properties of enormous graphs using just the eigenvectors associated with the Top-K largest eigenvalues.

In this work, we propose a hardware-optimized algorithm to approximate a solution to the Top-K eigenproblem on sparse matrices representing large graph topologies.
We prototype our algorithm through a custom FPGA hardware design that exploits HBM, Systolic Architectures, and mixed-precision arithmetic.
We achieve a speedup of 6.22x compared to the highly optimized ARPACK library running on an 80-thread CPU, while keeping high accuracy and 49x better power efficiency. 

\end{abstract}


\IEEEpeerreviewmaketitle

\section{Introduction}\label{sec:intro}

\begin{figure}[t]
    \centering
    \includegraphics[width=1\columnwidth]{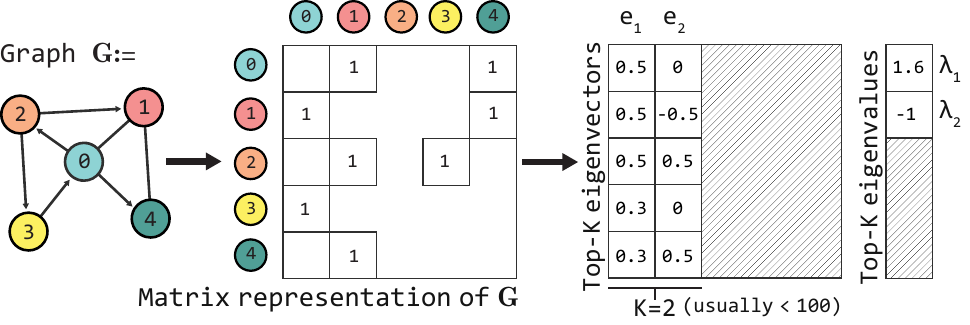}\\
    \caption{Top-K eigencomputation of a graph $G$, represented as a sparse matrix. While $G$ can have millions of vertices, we often need just the Top-K eigenvectors ($K=2$ in this example).}
    \label{fig:topk-eigencomputation}
\end{figure}

Research in information retrieval and recommender systems has spiked novel interest in spectral methods \cite{zhou2019SpectralRecommender}, a class of Machine Learning algorithms able to detect communities in large social and e-commerce graphs, and compute the similarity of graph elements such as users or products \cite{tang2020spectral}. 
At the core of many spectral methods lies the Top-K eigenproblem for large-scale sparse matrices, i.e. the computation of the eigenvectors associated with the largest eigenvalues (in modulo) of a matrix that stores only non-zero entries (\Cref{fig:topk-eigencomputation}). 
For example, the famous Spectral Clustering algorithm boils down to computing the largest eigenvalues of a sparse matrix representing the graph topology \cite{ng2002spectral}.
Despite the rise of theoretical interests for spectral methods, little research has focused on improving the performance and scalability of the Top-K sparse eigenproblem solvers, making them applicable to large-scale graphs.


Most existing high-performance implementations of eigenproblem algorithms operate on dense matrices and are completely unable to process matrices with millions of rows and columns (each encoding, for example, the user's friends in a social network graph) \cite{Myllykoski2020GPULIBQR}.
Even the highly optimized multi-core implementation of LAPACK requires more than 3 minutes to solve the full eigenproblem on a small graph with $\sim 10^4$ vertices and $\sim 50 \cdot 10^4$ edges on a Xeon 6248, as the eigenproblem complexity scales at least quadratically with the number of vertices in the graph.
Many implementations that support sparse matrices, on the other hand, are either forced to compute all the eigenvalues or require an expensive matrix inversion before solving the eigenproblem \cite{knyazev2001toward}.


The need for high-performance Top-K sparse eigenproblem algorithms goes hand in hand with custom hardware designs that can outperform traditional architectures in raw performance and power efficiency, given how applications on-top of Top-K eigenproblem are mostly encountered in data centers.
In this work, we tackle both problems by presenting a new Top-K sparse eigensolver whose building blocks are specifically optimized for high-performance hardware designs.

We introduce a novel algorithm to address the Top-K sparse eigenproblem and prototype it through a custom hardware design on an FPGA accelerator card; to the best of our knowledge this is the first FPGA-based Top-K sparse eigensolver.
Our algorithm is a 2-step procedure that combines the Lanczos algorithm (to reduce the problem size) \cite{lanczos1950iteration} with the Jacobi algorithm (to compute the final eigencomponents) \cite{Rutishauser1966jacobi}, as shown in \Cref{fig:fsfe_highlevel}. 
The Lanczos algorithm, often encountered in Top-K sparse eigensolvers \cite{nmeth2020scipy}, has never been combined with the Jacobi algorithm. 
Part of the reason lies in their different computational bottlenecks: the Lanczos algorithm demands large memory bandwidth, while the Jacobi algorithm is strongly compute-bound.
Our approach exploits the strengths of FPGA accelerator cards and overcomes the limitations of traditional architectures in this class of algorithms. 

First, the Lanczos algorithm presents a \gls{spmv} as its main bottleneck, an extremely memory-intensive computation with indirect and fully random memory accesses (\Cref{fig:fsfe_highlevel} \Circled{B}). 
Optimizing \gls{spmv} requires high peak memory bandwidth and fine-grained control over memory accesses, without passing through the traditional caching policies of general-purpose architectures.
Our hardware design features an iterative dataflow \gls{spmv} with multiple \glspl{cu}, leveraging every \gls{hbm} channels through a custom memory subsystem that efficiently handles indirect memory accesses.

Then, we introduce a \gls{sa} design for the Jacobi eigenvalue algorithm, a computationally-intensive operation that operates on reduced-size inputs ($K \times  K)$ (\Cref{fig:fsfe_highlevel} \Circled{D}). The Jacobi algorithm maps naturally to a \gls{sa} that ensures $\mathcal{O}(log(K))$ convergence, while traditional architectures do not ensure the same degree of performance. 
CPUs cannot guarantee that all the data are kept in L1 cache and are unlikely to have enough floating-point arithmetic units to parallelize the computation. This results in $\Omega(K^2)$ computational complexity and execution times more than 50 times higher than a \acrshort{fpga} (\Cref{sec:results}). Instead, GPUs cannot fill all their Stream Processors, as the input size is much smaller than what is required to utilize the GPU parallelism fully \cite{cosnau2014gpueigen}.

Moreover, our FPGA-based hardware design employs highly-optimized mixed-precision arithmetic, partially replacing traditional floating-point computations with faster fixed-precision arithmetic. While high numerical accuracy is usually demanded in eigenproblem algorithms, we employ fixed-precision arithmetic in parts of the design that are not critical to the overall accuracy and resort to floating-point arithmetic when required to guarantee precise results. 

In summary, we present the following contributions: 
\begin{itemize}
    \item A novel algorithm for approximate resolution of large-scale Top-K sparse eigenproblems (\Cref{sec:problem}), optimized for custom hardware designs. 
    \item A modular mixed-precision \acrshort{fpga} design for our algorithm that efficiently exploits the available programmable logic and the bandwidth of DDR and HBM (\Cref{sec:implementation}).
    \item A performance evaluation of our Top-K eigendecomposition algorithm against the multi-core ARPACK CPU library, showing a speedup of 6.22x and a power efficiency gain of 49x, with a reconstruction error due to mixed-precision arithmetic as good as $10^{-3}$ (\Cref{sec:results}).
\end{itemize}

\section{Related Work}\label{sec:soa}

To the best of our knowledge, no prior work optimizes Top-K sparse eigenproblem with custom FPGA hardware designs.

The most well-known large-scale Top-K sparse eigenproblem solver on CPU is the ARPACK library \cite{lehoucq1998arpack}, a multi-core Fortran library that is also available in SciPy and MATLAB through thin software wrappers. ARPACK implements the \gls{iram}, a variation of the Lanczos algorithm that supports non-Hermitian matrices.
Other sparse eigensolvers provide techniques optimized for specific domains or matrix types, although none is as common as ARPACK \cite{aktulga2012topology,hernandez2009survey,maschhoff1996p_arpack,lee2018solution}.

On GPUs, the cuSOLVER \cite{cusolver} library by Nvidia provides a simple eigensolver based on the shift-inverse method that retrieves only the largest eigenvalue and its eigenvector (i.e. $K=1$), which is significantly more limited than the general Top-K eigenproblem.
The nvGRAPH library \cite{nvgraph}, also developed by Nvidia, provides an implementation of spectral clustering at whose core lies the Lanczos algorithm. However, the implementation of the inner Lanczos algorithm is not publicly available.
To the best of our knowledge, there is no publicly available GPU implementation of the Lanczos algorithm that can solve large scale sparse eigenproblems required by spectral methods.
The MAGMA library \cite{magma} solves the Top-K sparse eigenproblem through the alternative LOBPCG algorithm \cite{knyazev2001toward}, which requires multiple iterations (each containing at least one \gls{spmv}) to compute even a single eigenvector, to the contrary of the Lanczos algorithm.
Other GPU-based Top-K eigensolvers are domain-specific, do not support large-scale inputs, or do not leverage features of modern GPUs such as HBM memory or mixed-precision arithmetic \cite{evstigneev2017implementation,dubois2011accelerating}. 
Eigensolvers for dense matrices are more common on GPUs,
as they easily exploit the enormous memory bandwidth of these architectures:
Myllykoski et al. \cite{Myllykoski2020GPULIBQR} focus on accelerating the case of dense high-dimensional matrices (around $10^5$ rows) while Cosnuau \cite{cosnau2014gpueigen} operates on multiple small input matrices. Clearly, none of the techniques that operate on dense matrices can easily scale to matrices with millions of rows as simply storing them requires terabytes of memory.

Specialized hardware designs for eigensolvers are limited to resolving the full eigenproblem on small dense matrices, through the QR-Householder Decomposition and Jacobi eigenvalue algorithm. 
Most formulations of the Jacobi algorithm \cite{gupta2019eigen,Burger2020Embedded} leverage \acrlong{sa}, a major building block of high performance domain-specific architectures from their inception \cite{kung1982systolic} to more recent results \cite{jouppi2017datacenter,langhammer2020high,asgari2020proposing}.
However, hardware designs of the Jacobi algorithm based on \gls{sa} cannot scale to large matrices, as the resource utilization scales linearly with the size of the matrix.
Implementations of the \textit{QR-Houseolder} algorithm face similar problems \cite{vazquez2020fpga,aslan2012qr} as they also leverage systolic architectures, although research research about resource-efficient designs do exist \cite{Desai2017QRHLS}. 
\section{Solving the Top-K Sparse Eigenproblem}\label{sec:problem}

\begin{figure}[t]
    \centering
    \includegraphics[width=1\columnwidth]{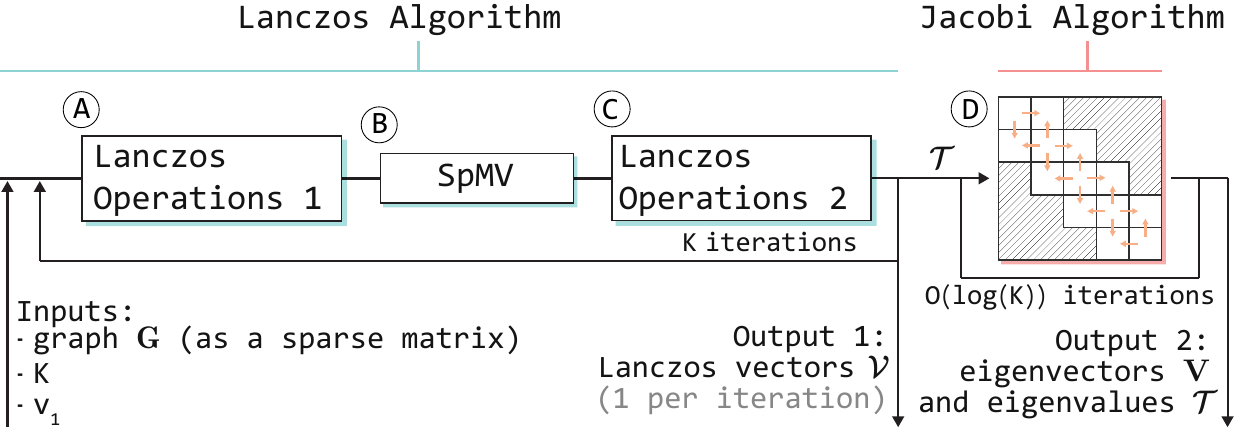}\\
    \caption{Steps of our novel Top-K sparse eigenproblem solver, which combines the Lanczos algorithm with a Systolic Array formulation for the Jacobi eigenvalue algorithm.}
    \label{fig:fsfe_highlevel}
\end{figure}

Algorithms like Spectral Clustering contain as their core step a Top-K sparse eigenproblem, i.e. finding eigenvalues and eigenvectors of sparse matrices representing, for instance, graph topologies with millions of vertices and edges.

\algrenewcommand\algorithmicindent{1.2em}%
\begin{algorithm}[t]
  \caption{Lanczos algorithm for the Top-K eigenvectors}\label{alg:lanczos}
  \begin{algorithmic}[1]
    \Function{Lanczos}{$\mathbf{M},K,v_1$} 
    \State $\beta_1 \gets 0, v_0 \gets \textbf{0}_N$ \Comment{Initialization}

    \For{$i$ in $1, K$}     \Comment{\textbf{Main Lanczos loop}}
    \If{$i > 1$}
    \State $\beta_i \gets$ $\|w'_{i - 1}\|_2$
    \State $v_i \gets w'_{i - 1} / \beta_i$ \Comment{Compute new Lanczos vector}
    \EndIf
    \State $w_i \gets \mathbf{M}v_i$ \Comment{Sparse matrix-vector multiplication}
    \State $\alpha_i \gets w_i v_i$
    \State $w'_i \gets w_i - \alpha_i v_{i}  - \beta_i v_{i - 1}$
    \State Orthogonalize $w'_i$ with respect to $\mathcal{V}$
    \EndFor
    \LineComment{\textbf{Tridiagonal matrix $\mathcal{T}$ and Lanczos vectors $\mathcal{V}$}}
    \State Output $\{\mathcal{T} = [\alpha_1, \ldots, \alpha_K], [\beta_1, \ldots, \beta_{K - 1}]\}$ 
    \State Output $\mathcal{V} = [v_1, \ldots, v_K]$
    
    \EndFunction
  \end{algorithmic}
\end{algorithm}
\begin{figure}[b]
\centering
     $
        \begin{bmatrix}
            \alpha_1 & \beta_1  & 0        & 0        & 0       \\
            \beta_1  & \alpha_2 & \beta_2  & 0        & 0       \\
            0        & \beta_2  & \alpha_3 & \beta_3  & 0       \\
            0        & 0        & \beta_3  & \alpha_4 & \beta_4 \\ 
            0        & 0        & 0        & \beta_4  & \alpha_5
        \end{bmatrix}
    $ 
    \caption{Example of ($5 \times 5$) tridiagonal matrix, obtained as output of the Lanczos algorithm for $K=5$.}
    \label{fig:tridiag}
    
\end{figure}

Given a sparse square matrix $\mathbf{M} \in \mathbb{R}^{n\times n}$ and an integer $K \ll n$ the goal of the Top-K sparse eigenproblem is to find the $K$ eigenvalues with the highest magnitude, and their associated eigenvectors.
This is equivalent to computing the approximate decomposition $\mathbf{M}\!\approx\!\mathbf{M}_K\!=\!\mathbf{Q}_K \mathbf{\Lambda}_K \mathbf{Q}_K^T$, with $\mathbf{Q}_K \in \mathbb{R}^{n\times K}$ and $\mathbf{\Lambda}_K \in \mathbb{R}^{K\times K}$.
$\mathbf{Q}_K$ contains the eigenvectors, while $\mathbf{\Lambda}_K$ is a diagonal matrix containing the eigenvalues.
Indeed, computing all the $n$ eigenvalues of the matrix is intractable for large matrices and redundant for many applications that require only a handful of eigencomponents.
For example, Spectral Clustering and many of its variations rely only on the Top-K eigenvectors, with K rarely above $\sim 10$. 

In this work, we propose a novel algorithm to solve the Top-K sparse eigenproblem, combining the Lanczos algorithm and the Jacobi eigenvalue algorithm. Our technique is particularly suited for highly optimized and modular hardware designs.
The first phase leverages the Lanczos algorithm, taking as input the original matrix $\mathbf{M}$, the number of desired eigencomponents $K$ and an \textit{L2-normalized} random vector $v_1 \in \mathbb{R}^{n}$, initialized with values equal to $1/n^2$.
The Lanczos algorithm outputs a $K \times K$ symmetric tridiagonal matrix $\mathcal{T}$ (\Cref{fig:tridiag}) and a set of orthogonal Lanczos vectors $\mathcal{V} \in \mathbb{R}^{K \times n}$.
As second step, we apply the Jacobi eigenvalue algorithm to $\mathcal{T}$.
This algorithm transforms $\mathcal{T}$ into a diagonal matrix containing its eigenvalues, and returns a matrix $\mathbf{V}$ with the eigenvectors of $\mathcal{T}$.
Each eigenvalue $\lambda$ of $\mathcal{T}$ is also an eigenvalue of the original matrix $\mathbf{M}$. Moreover, if $x$ is the eigenvector of  $\mathcal{T}$ associated to $\lambda$, then $\mathcal{V}x$ is the eigenvector of $\mathbf{M}$ associated to $\lambda$.
$\mathbf{M}_K$ can be obtained as $\mathbf{M}_K\!=\! (\mathcal{V}\mathbf{V})\mathcal{T}(\mathcal{V}\mathbf{V})^T$, although many applications in spectral analysis only require the Top-K eigenvalues and eigenvectors of $\mathbf{M}$ instead of retrieving $\mathbf{M}_K$.


\subsection{The Lanczos Algorithm}

The Lanczos algorithm retrieves the Top-K eigencomponents of a matrix and is often employed as a building block of large-scale eigenproblem algorithms \cite{lehoucq1998arpack,golub1977block,calvetti1994implicitly}.
The $K \times K$ output tridiagonal matrix $\mathcal{T}$ is significantly smaller than the input ($K \ll n$) and also simpler in structure, as elements outside of the band enclosing the main diagonal and the ones immediately above and below are zero. 
Pseudo-code of the algorithm is provided in \Cref{alg:lanczos}.
For each of the $K$ iterations, it computes a Lanczos vector $v_i$ by normalizing $w'_{i-1}$, obtained at the previous iteration (line 6 and \Cref{fig:fsfe_highlevel}A).
From $v_i$, we obtain $w'_i$ by projecting the matrix $\textbf{M}$ into $v_i$ (line 7 and \Cref{fig:fsfe_highlevel}B), followed by an orthogonalization (lines 8 to 10 and \Cref{fig:fsfe_highlevel}C).
The algorithm is highly efficient as each vector $v_i$ is computed in a single iteration, and $K \ll n$.

\algrenewcommand\algorithmicindent{1.2em}%
\begin{algorithm}[t]
  \caption{Jacobi eigenvalue algorithm with Systolic Arrays}\label{alg:jacobi}
  \begin{algorithmic}[1]
   \Function{Jacobi}{$\mathcal{T}$}
    \State $\mathbf{V} \gets \mathbb{1}_K$ \Comment{Identity matrix of size $K \times K$}
    \Repeat
        \For{$i $ in $ 1, K / 2$} \Comment{Diagonal \gls{cu}}
            \State $p_{ii} \gets \mathcal{T}[2i:2i + 1, 2i:2i + 1]$
            \State $\theta_i \gets \frac{1}{2}\arctan{\frac{2\beta}{\alpha - \delta}}$
            \State Rotate $p_{ii}$ \Comment{\textbf{Full equation in \Cref{fig:diagonalproc}}}
            \State Propagate $c_i$ and $ s_i$
        \EndFor
        \For{$j$ in $1, K / 2 - 2$} \Comment{Offdiagonal \gls{cu}}
            \State $i \gets j + 1$
            \State Receive $c_i, c_j$, $s_i, s_j$ from $p_{ii}, p_{jj}$
            \State $p_{ij} \gets  \mathcal{T}[2i:2i + 1, 2j:2j + 1]$
            \State Rotate $p_{ij}$ \Comment{\textbf{Full equation in \Cref{fig:offdiagonalproc}}}
        \EndFor
        \For{$i$ in $1, K / 2$} \Comment{Eigenvector \gls{cu}}
            \For{$j$ in $1, K / 2$}
            \State $v_{ij} \gets \mathbf{V}[2i:2i + 1, 2j:2j + 1]$
            \State Receive $c_j, s_j$ from $p_{jj}$
            \State Rotate $v_{ij}$ \Comment{\textbf{Full equation in \Cref{fig:eigenvector_processor}}}
            \EndFor
        \EndFor
        \State Permute rows and columns of $\mathcal{T}$ and $\mathbf{V}$ \Comment{\Cref{fig:jacobi}}
    \Until $\mathcal{T}$ becomes diagonal
    \State Output $\mathcal{T}$ \Comment{Eigenvalues of the input $\mathcal{T}$}
    \State Output $\mathbf{V}$ \Comment{Eigenvectors of the input $\mathcal{T}$}
    \EndFunction
  \end{algorithmic}
\end{algorithm}
\begin{figure}
    \centering
    \begin{subfigure}{\columnwidth}
        \begin{center}
        $
        \begin{bmatrix}
            c_{i}  & s_{i}   \\
            -s_{i}  & c_{i}  \\
        \end{bmatrix}
        \begin{bmatrix}
            \alpha  & \beta     \\
            \gamma  & \delta \\
        \end{bmatrix}
        \begin{bmatrix}
            c_{i}  & -s_{i}   \\
            s_{i}  & c_{i}  \\
        \end{bmatrix}
         = 
        \begin{bmatrix}
            \alpha'     & 0    \\
            0  & \delta'    \\
        \end{bmatrix}
        $
        \caption{Operations for the Diagonal Processor $p_{ii}$ (\Cref{fig:jacobi}A).}
        \label{fig:diagonalproc}
        \end{center}
    \end{subfigure}\vspace{3mm}
    \begin{subfigure}{\columnwidth}
        \begin{center}
        $
        \begin{bmatrix}
            c_{i}  & s_{i}   \\
            -s_{i}  & c_{i}  \\
        \end{bmatrix}
        \begin{bmatrix}
            \alpha  & \beta     \\
            \gamma  & \delta \\
        \end{bmatrix}
        \begin{bmatrix}
            c_{j}  & -s_{j}   \\
            s_{j}  & c_{j}  \\
        \end{bmatrix}
         = 
        \begin{bmatrix}
            \alpha' & \beta'\\
            \gamma' & \delta'\\
        \end{bmatrix}
        $
        \caption{Operations for the Offdiagonal Processor $p_{ij}$ (\Cref{fig:jacobi}C).}
        \label{fig:offdiagonalproc}
        \end{center}
    \end{subfigure}\vspace{3mm}
    \begin{subfigure}{\columnwidth}
        \begin{center}
        $
        \begin{bmatrix}
            w  & x     \\
            y  & z \\
        \end{bmatrix}
        \begin{bmatrix}
            c_{j}  & -s_{j}   \\
            s_{j}  & c_{j}  \\
        \end{bmatrix}
         = 
        \begin{bmatrix}
            w'     & x'    \\
            y'  & z'   \\
        \end{bmatrix}
        $
        \caption{Operations for the Eigenvector Processor $p_{ij}$ (\Cref{fig:jacobi}D).}
        \label{fig:eigenvector_processor}
        \end{center}
    \end{subfigure}
    \caption{Operations performed by different processors in the Jacobi eigenvalue Systolic Array architecture. Values $c_{i}$ and $s_{i}$ indicate $cos(\theta_i)$ and $sin(\theta_i)$, with $\theta_i = \frac{1}{2}\arctan{\frac{2\beta}{\alpha - \delta}}$.}
    \label{fig:jacobiequations}
\end{figure}

The Lanczos algorithm is particularly efficient on sparse matrices, as its most expensive operation is an iterative \gls{spmv}, bounding its computational complexity to $\mathcal{O}(K \cdot E)$, with $E$ being the number of non zero elements of $\mathbf{M}$.
In our hardware design, we optimize the memory-intensive \gls{spmv} computation through multiple independent \glspl{cu}, so that we can take advantage of all the available 32 HBM channels of a Xilinx Alveo U280 FPGA accelerator card (\Cref{sec:spmvarch}).

This algorithm is prone to numerical instability as the Lanczos vectors $\mathcal{V}$ can quickly lose pairwise orthogonality if $K$ is very large.
To prevent instability, we normalize the input matrix in \textit{Frobenius norm} as eigencomponents are invariant to constant scaling:
values of the matrix are in the range $(-1, 1)$, which implies that eigenvalues and eigenvectors are also in the range $(-1, 1)$. This property enables the use of fixed-point arithmetic to improve performance and reduce resource usage (\Cref{sec:accuracy}).
We further improve numerical stability by adopting a version of the algorithm that reorders operations \cite{paige1972computational} and reorthogonalizes Lanczos vectors in each iteration \cite{parlett98}.
Reorthogonalization (\Cref{alg:lanczos}, line 10) requires $K^2/2$ more operations of cost $\mathcal{O}(n)$, increasing complexity to $\mathcal{O}(K (E + nK^2/2))$. 
We also introduce the option of performing reorthogonalization every 2 iterations, for a lower overhead of $\mathcal{O}(n(K/2)^2/2)$, with negligible accuracy loss (\Cref{sec:accuracy}).
In practice, execution time is usually dominated by \gls{spmv} making reorthogonalization a viable option.

\subsection{The Jacobi Eigenvalue Algorithm}

The Jacobi eigenvalue algorithm computes the eigenvalues and eigenvectors of a dense symmetric real matrix. It is an iterative procedure that performs rotations on square submatrices.
Each iteration is highly computationally-intensive as it contains $\Omega(K^2)$ trigonometric operations. 
However, this algorithm is particularly well suited to solve eigenproblems on small tridiagonal matrices. As many matrix values are zero and cannot introduce data-dependencies in rotations, it is possible to parallelize the entire computation at hardware-level.

\begin{figure}[t]
    \centering
    \includegraphics[width=1\columnwidth]{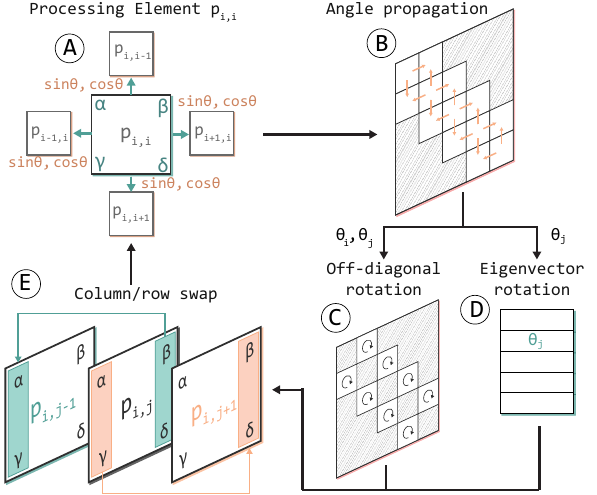}\\
    \caption{Steps of the Jacobi eigenvalues computation using Systolic Arrays. Each \acrfull{pe} $p_{ij}$ holds 4 values $\alpha$, $\beta$,  $\gamma$, $\delta$, and $\theta = \frac{1}{2}\arctan{\frac{2\beta}{\alpha - \delta}}$.}
    \label{fig:jacobi}
\end{figure}

The Jacobi eigenvalue algorithm has sought many formulations to improve either its parallelism or its resource utilization.
The best-known formulation of the algorithm was proposed by Brent and Luk \cite{brent1984solution} and has been the standard for implementing the algorithm on \acrshort{fpga} to this day \cite{gupta2019eigen, Burger2020Embedded}. Our design improves this formulation with a more resource-efficient procedure for interchanging rows and columns, and its structure is shown in \Cref{fig:jacobi}.

We employ a \gls{sa} design that maps the input matrix as $2 \times 2$ submatrices to $K^2/4$ adjacent \textit{processors} (or \gls{cu}) (\Cref{fig:jacobi}, \Circled{A}).
The systolic architecture propagates the rotation angles \Circled{B} and the values stored in each processors \Circled{E}.

Starting from $\mathcal{T}$, the algorithm set to zero $K / 2$ off-diagonal entries per iteration by using rotations.
Diagonal processors annihilate $\beta$ and $\gamma$ components (\Cref{alg:jacobi}, line 7) with a rotation of angle $\theta$.
This angle is propagated (line 8) to the off-diagonal processor (line 9), and to the eigenvector processor that applies the same rotation to the identity matrix (line 14).

To ensure convergence, diagonal \glspl{cu} are fed non-zero elements at each iteration in the $\beta$ and $\gamma$ position. 
New non-zero elements are provided to the diagonal \glspl{cu} by swapping rows and columns, since eigencomponents are invariant to linear combinations.
We improve the swap procedures of Brent and Luk  \cite{brent1984solution} by swapping vectors \textit{in reverse}, obtaining the same results with fewer resources (\Cref{sec:col_row_swap}).

The \gls{sa} formulation allows performing each iteration of the algorithm in constant time, enabling complexity equal to the number of iterations, $\mathcal{O}(log(K))$, instead of having cost above $\Omega(K^{2}\cdot \log(K))$ due to the matrix multiplications \cite{brent1984solution}. 

\section{The Proposed Hardware Design}\label{sec:implementation}

\begin{figure}[t]
    \centering
    \includegraphics[width=1\columnwidth]{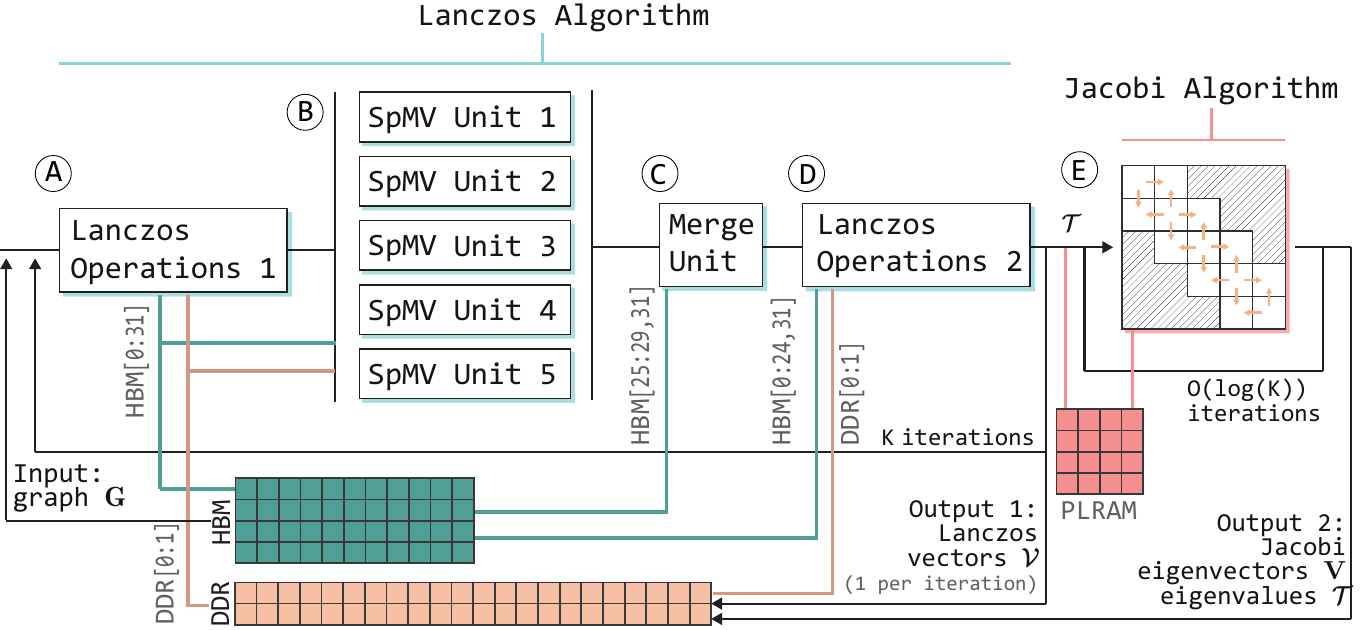}\\
    \caption{High-level architecture of our Top-K Sparse Eigencomputation FPGA design. We highlight interconnections between FPGA computational units and the FPGA board memory.}
    \label{fig:fsfe}
\end{figure}

This section presents our custom FPGA-based hardware design for the Top-K sparse eigenproblem algorithm previously introduced.
The logical division between the Lanczos and Jacobi algorithms is also present in the hardware implementation.
Our hardware design is composed of two macro-areas that are mapped to separate reconfigurable \glspl{slr} of the FPGA, to provide more efficient resource utilization and higher flexibility in terms of clock frequency, memory interfaces, and reconfigurability.
\Cref{fig:fsfe} shows a high level view of our \acrshort{fpga} design.
We prototyped our hardware design on an Alveo U280 accelerator card with HBM2 and DDR4 memory.
The Lanczos algorithm, being a memory-intensive computation, is mapped to SLR0, which provides direct access to all the HBM2 memory interfaces on the accelerator card. 
SLR1 and SLR2 hosts different replicas of the IP core implementing the Jacobi algorithm, optimized for different numbers of eigenvectors $K$.

\subsection{Lanczos Hardware Design}\label{sec:Lanczosarch}

The left part of \Cref{fig:fsfe} highlights the Lanczos algorithm hardware design components.
Partitions of the sparse input matrix are read from \gls{hbm} \Circled{A} and distributed to 5 parallel \gls{spmv} \glspl{cu} \Circled{B} (\Cref{alg:lanczos}, line 7). Partial results from every partition are merged \Circled{C} into a single vector to be used by the remaining linear operations \Circled{D} (lines 5, 6, 8, 9). Operations are then repeated $K$ times to produce the $3 \cdot K - 2$ values in the tridiagonal matrix $\mathcal{T}$ and the $K$ Lanczos vectors in $\mathcal{V}$, stored in DDR memory. 

\subsection{SpMV Hardware Design}\label{sec:spmvarch}

The biggest bottleneck in the Lanczos algorithm is an iterative \gls{spmv} computation (\Cref{alg:lanczos}, line 7). While other computations in the Lanczos algorithm are relatively straightforward to optimize and parallelize, \gls{spmv} is well-known for being a complex, memory-intensive computation that presents indirect and random memory accesses \cite{nguyen2020FPGAPotential}. 
Although significant research has been made into developing high-performance \gls{spmv} implementations on FPGA \cite{10.1145/3352460.3358330,grigoras2015accelerating,umuroglu2015vector,parravicini2021scaling,jain2020domain}, the Lanczos algorithm introduces circumstances that prevents us from using an out-of-the-box FPGA \gls{spmv} implementation.
Our \gls{spmv} design must perform multiple iterations without communication from device to host, as data-transfer and synchronizations would hinder performance.
Then, the \gls{spmv} must be easily partitioned and replicated to provide flexibility over the hardware resources.
Finally, we require access to multiple HBM channels to maximize the overall memory bandwidth achieved in the computation.

Our final \gls{spmv} design extends and improves the one recently proposed by Parravicini et al. \cite{parravicini2020reduced} in the context of graph ranking algorithms, which are also variations of the power iteration method as in the case of the Lanczos algorithm.
Below we introduce how we leveraged HBM memory in our \gls{spmv} design to provide better scalability and performance.

\begin{figure}[t]
    \centering
    \includegraphics[width=1\columnwidth]{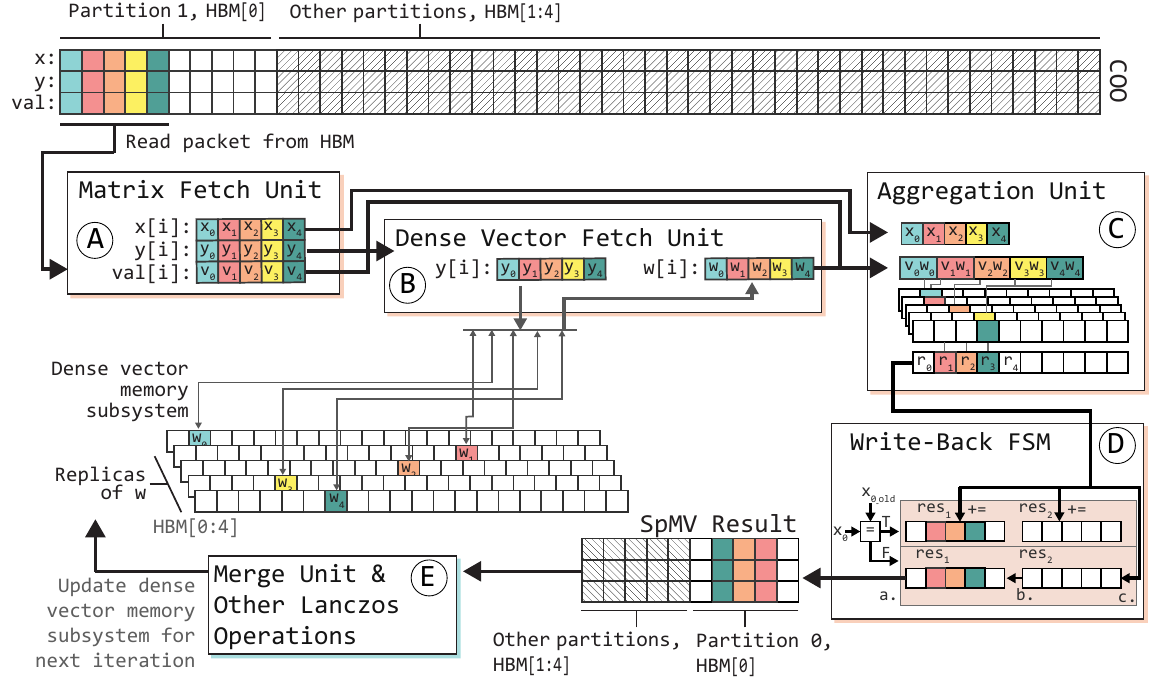}\\
    \caption{Architecture of one iterative \acrshort{spmv} \acrshort{cu}.
    Each \acrshort{cu} processes a portion of the input matrix through a 4-stage dataflow design, and results are replicated on the \textit{dense vector memory subsystem} after each iteration.
    }
    \label{fig:spmv}
\end{figure}

\subsubsection{SpMV Dataflow Architecture}\label{sec:spmvdataflow}

\begin{figure}[t]
    \centering
    \includegraphics[width=1\columnwidth]{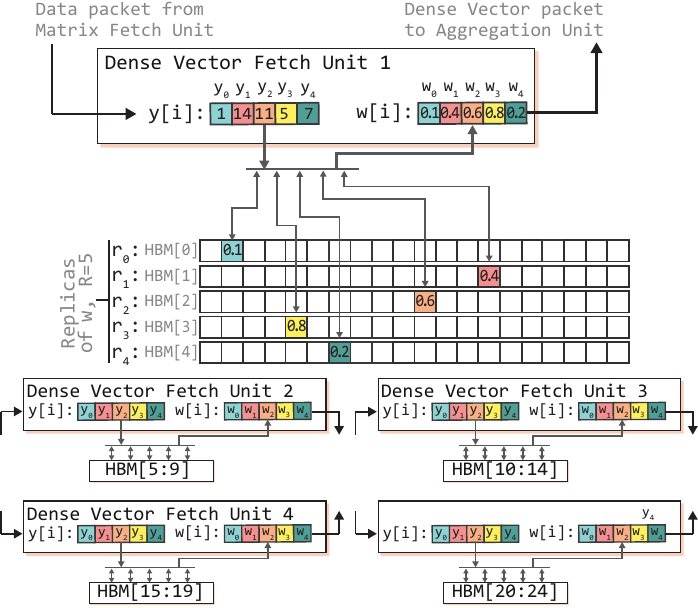}\\
        \caption{Dense vector memory subsystem of our \gls{spmv} FPGA design. Index $y_i$ accesses replica $r_i$, guaranteeing a pipelined design with 5 random vector accesses per clock cycle.}
    \label{fig:spmvsubsystem}
\end{figure}

As \gls{spmv} is an extremely memory-intensive computation, a good \gls{spmv} implementation should make efficient use of the memory bandwidth made available by the underlying hardware. 
\Cref{fig:spmv} shows the structure of one of our \gls{spmv} \glspl{cu}.
We employ a streaming dataflow \gls{spmv} design that reads the input sparse matrix stored using the \gls{coo} format. 
In the \gls{coo} format, non-zero entries of the matrix are stored using 3 32-bits values: the row and column index in the matrix and the value itself.
Compared to other sparse matrix data-layouts, such as \gls{csr}, the COO format does not present indirect data accesses that can severely reduce the opportunities for a pipelined design. 
The \textit{Matrix Fetch Unit} in each \gls{cu} is connected to a single HBM channel and reads, for each clock cycle, a packet of 512 bits containing 5 non-zero matrix entries \Circled{A}. 
Memory transactions happen in continuous bursts of maximum AXI4 length (256 beats): each \gls{cu} reads the matrix at the maximum bandwidth offered by the HBM channel (14.37 GB/s, for a total of 71.87 GB/s using 5 \gls{cu}).
For each of the 5 non-zero values in each COO packet, the \textit{Dense Vector Fetch Unit} performs a random access to the \gls{spmv} dense vector \Circled{B}. This step is critical to the overall \gls{spmv} performance: compared to \cite{parravicini2020reduced}, we leverage HBM instead of \gls{uram}, achieving better scalability and performance. We detail our \textit{Dense Vector Memory Subsystem} below and in \Cref{fig:spmvsubsystem}.
The \textit{Aggregation Unit} sums results within a single data-packet that refers to the same matrix column \Circled{C}.
A \textit{Write-Back Finite-State Machine} stores results of each \gls{cu} to HBM \Circled{D}. Each write-transaction is a 512-bits data-packet containing up to 15 values, each referring to a single matrix row. Compared to \cite{parravicini2020reduced} we reduce the number of write transactions by 3 times the average number of non-zeros per row. As such, we can store results through the same HBM channels of the dense vector with no detriment to performance.

Compared to the original \gls{spmv} design in \cite{parravicini2020reduced}, we support multiple \gls{spmv} \glspl{cu} that operate on partitions on the \gls{coo} input matrix, created by assigning an equal number of rows to each \gls{cu}.
We employ up to 5 \gls{spmv} \glspl{cu} (\Cref{fig:fsfe}). 
While in principle it is possible to place more \glspl{cu}, our current design is limited by the hardened AXI switch in the Alveo U280 that prevents the use of more than 32 AXI master channels to HBM, which we fully employ \cite{alveohbmcontroller}. 
Each \gls{spmv} \gls{cu} compute a portion of the output vector: partial results are aggregated by the \textit{Merge Unit} (\Cref{fig:fsfe} \Circled{C}) and replicated across HBM channels to use them in the following iteration.

\subsubsection{SpMV Dense Vector Memory Subsystem}\label{sec:spmvsubsystem}

Each \gls{spmv} \gls{cu} processes 5 non-zero matrix entries per clock cycle, and for each non-zero entry it must perform a random access on a dense vector of size $n$ (in our case, the Lanczos vector $v_i$ at iteration $i$). 
As each AXI master channel can handle only one read transaction per cycle, we need to replicate the dense vector 5 times, similarly to \cite{kestur2012towards}. 
The hardened AXI switch in the Alveo U280 renders highly inefficient to attach multiple AXI master channels to the same HBM bank: only 32 AXI master channels are available, and small memory transactions (32 bits) have the same performance as larger transactions, preventing sustained bandwidth sharing \cite{choi2020hls,lu2021demystifying,wang2020shuhai}.
We solve the issue by leveraging the abundant HBM memory on the Alveo U280, and replicating the dense vector 5 times for each \gls{cu}, as in \Cref{fig:spmvsubsystem}.
A more flexible AXI switch would enable multiple 32-bits read transactions on the same HBM channel in a single clock cycle, reducing the demand for data replication.
Compared to \cite{parravicini2020reduced}, our HBM-based memory subsystem marks a significant improvement, as we avoid \gls{uram} to store the intermediate dense vector and results.
Instead of being limited by the FPGA's 90 MB of \gls{uram}, we store the dense vector using individual HBM banks with 250 MB of capacity, allowing computations on matrices with up to
62.4 million rows.
Moreover, high \gls{uram} consumption significantly limits the maximum attainable frequency, while we do not incur in this limitation (\Cref{tab:resources}).

\subsection{Jacobi Systolic Array Design}\label{sec:jacobiarch}

The Jacobi eigenvalue algorithm is very computationally intensive. Although it processes a small input of size $K \times K$, unoptimized implementations still require a significant amount of time due to a large number of dense matrix multiplications. Moreover, its convergence rate is implementation-dependent and as high as $\mathcal{O}(K^2)$.
By adopting a \gls{sa}-based design, we overcome both issues.
By parallelizing the computation through a \gls{sa} formulation and performing rotations concurrently, we decrease the number of iterations for convergence to $\mathcal{O}(log(K))$.
Rotations,  equivalent to multiplications on $2 \times 2$ submatrices, are unrolled and performed in constant time.

Our design for the Jacobi algorithm is optimized to compute up to $K$ eigenvalues. While it can compute a lower amount of eigenvalues without a reconfiguration, we place in the same FPGA bitstream multiple \textit{Jacobi cores} optimized for specific $K$ (4, 8, 16, etc.).
We can configure both SLR1 and SLR2 with Jacobi cores to fully utilize the FPGA resources and opening the doors for independent optimization on specific values of $K$ by reconfiguring individual \glspl{slr}.
The \textit{Lanczos Core} on \gls{slr} transfers only the $3K - 2$ values of $\mathcal{T}$ to the Jacobi cores on \gls{slr}1 and \gls{slr}2.
We prevent inefficiencies related to inter-\gls{slr} communication by moving data through PLRAM, while also avoiding the long read-write latency of DDR and HBM.

In practice, the systolic formulation cannot scale beyond very small matrices ($K \approx 32$) due to the large number of resources required for trigonometric operations in each \gls{cu}. 
While resource utilization has prevented widespread adoption of the Jacobi algorithm for general eigenproblem resolution, it is not a limitation for our use case, as we apply the Jacobi eigenvalue algorithm on small $K \times K$ inputs by design.

On CPU, approaches such as QR factorization are more common \cite{buttari2008parallel}, because efficient systolic array formulations of the Jacobi algorithm require full control over cache eviction policies.
Moreover, even modern CPUs lack enough floating-point arithmetic units to perform the operations required for an iteration at once:
even for a small $K$ such as $K = 8$, the Jacobi algorithm computes 16 trigonometric operations and about $800$ floating-point multiplications per iteration.

Instead, we leverage the abundant hardware resources of our \acrshort{fpga} platform to perform all these operations concurrently, making it the optimal choice for our Jacobi \gls{sa} design.



\subsubsection{Diagonal And Offdiagonal \gls{cu}}\label{sec:diag_offdiag}
Diagonal \gls{cu} (\Cref{alg:jacobi}, line 4) annihilate elements immediately outside the diagonal via a matrix rotation. Although the rotation angle is arbitrary, the fastest convergence is achieved by setting $\theta = \frac{1}{2}\arctan{\frac{2\beta}{\alpha - \delta}}$, which eliminates the $\beta$ and $\gamma$ components (\Cref{fig:diagonalproc}).
We efficiently compute the components of the rotation matrix via Taylor series expansion. Even an order-3 approximation provides excellent accuracy ($\sim 10^{-6}$ at $\pm \frac{\pi}{4}$), using significantly fewer DSPs and BRAMs than the CORDIC core.
Rotation on the diagonal (\Cref{fig:jacobi} \Circled{A}) are performed by $K / 2$ parallel cores, propagating rotation values \Circled{B} in constant time to the Offdiagonal \gls{cu} (\Cref{alg:jacobi}, line 9). 
As each \glspl{cu} holds only 4 elements, matrix multiplications are fully unrolled and performed in constant time.
Eigenvectors (\Cref{alg:jacobi}, line 14) \Circled{D} are computed in parallel to the rotation of the Offdiagonal \gls{cu} \Circled{C} as they only require rotation values.

\subsubsection{Row/Column Interchange}\label{sec:col_row_swap}
Each \gls{cu} has 8 connections to propagate input and output values of $\alpha, \beta, \gamma, \delta$ values to adjacent processors, in addition to communicating the rotation value $\theta$.
As shown in \Cref{fig:jacobi}E, each processor $p_{i,j}$ with $i$ and $j \neq (1, K / 2)$ propagates its $\alpha$ and $\gamma$ values to the $\beta$ and $\delta$ slots of $p_{i, j + 1}$ and its $\beta$ and $\delta$ values to the $\alpha$ and $\gamma$ slots of $p_{i, j - 1}$.
Processors in the first column ($p_{i,1}$) propagate $\beta$ and $\delta$ to to the $\alpha$ and $\gamma$ slots of $p_{i, 2}$.
Processors $p_{i, K / 2}$ propagate $\beta$ and $\delta$ to their own $\alpha$ and $\gamma$ slots.
Operations for the column interchange are symmetrical.
As $\alpha$ and $\gamma$ of $p_{i, 1}$ are never propagated, more swaps are performed towards lower indices than higher indices. These additional swaps require $K$ temporary vectors to store rows that would be overwritten by the swaps.
To avoid wasting resources for these temporary vectors, we execute operations in \textit{reverse}, from $K / 2$ to $1$.
As row/column swaps do not introduce additional data dependencies, we perform them in a single clock cycle using FFs.

\section{Experimental Evaluation}\label{sec:results}

\renewcommand\theadalign{tl}
\renewcommand\theadfont{\bfseries}
\setlength\tabcolsep{3pt}

\begin{table}
\centering
\ra{1.2}
    \caption{Resource usage and clock frequency in our FPGA hardware design, divided by algorithm.
    }\label{tab:resources}
    \resizebox{1\linewidth}{!}{
	\begin{tabular}{@{}lllllllll@{}}
		\toprule
    	\thead{Algorithm} & \phantom{abc} & \thead{SLR} &
    	\thead{LUT} & \thead{FF} & \thead{BRAM} & \thead{URAM} & \thead{DSP} & \thead{Clock (MHz)}\\
	   
	    \midrule
    	\textbf{Lanczos} && SLR0 & 42\%  & 13\% & 15\% & 0\% & 16\% & 225 \\
    	\textbf{Jacobi}  && SLR1 & 40\% & 42\% & 0\%  & 0\% & 68\% & 225 \\   
    	\textbf{Jacobi}  && SLR2 & 15\% & 17\% & 0\%  & 0\% & 34\% & 225 \\   
    	\midrule
        \textbf{Available} &&& 1097419 & 2180971 & 1812 & 960 & 9020 &  \\
		\bottomrule
	\end{tabular}
    }
\end{table}
\renewcommand\theadalign{tl}
\renewcommand\theadfont{\bfseries}

\begin{table}
\centering
    \caption{Matrices/graphs in the evaluation, sorted by number of edges/non-zero entries (in millions). For each matrix, we report the memory footprint when stored as COO.}
    \resizebox{1\columnwidth}{!}{
	\begin{tabular}{@{}llllll@{}}
		\toprule
		\thead{ID} & \thead{Name} & \thead{Rows (M)}  & \thead{Non-zeros (M)} & \thead{Sparsity (\%)} & \thead{Size (GB)} \\
		\midrule
        \textbf{WB-TA} & wiki-Talk & 2.39 & 5.02 & $\SI{8.79e-04}{}$ & $\SI{0.06}{\giga\byte}$ \\
        \textbf{WB-GO} & web-Google & 0.91 & 5.11 & $\SI{6.17e-04}{}$ & $\SI{0.07}{\giga\byte}$ \\
        \textbf{WB-BE} & web-Berkstan & 0.69 & 7.60 & $\SI{1.60e-03}{}$ & $\SI{0.10}{\giga\byte}$ \\
		\textbf{FL} & Flickr & 0.82 & 9.84 & $\SI{1.46e-03}{}$ & $\SI{0.13}{\giga\byte}$ \\
		\textbf{IT} & italy\_osm & 6.69 & 14.02 & $\SI{3.13e-05}{}$ & $\SI{0.18}{\giga\byte}$ \\
		\textbf{PA} & patents & 3.77 & 14.97 & $\SI{1.05e-04}{}$ & $\SI{0.19}{\giga\byte}$ \\
		\textbf{VL3} & venturiLevel3 & 4.02 & 16.10 & $\SI{9.96e-05}{}$ & $\SI{0.21}{\giga\byte}$ \\
		\textbf{DE} & germany\_osm & 11.54 & 24.73 & $\SI{1.86e-05}{}$ & $\SI{0.32}{\giga\byte}$ \\
		\textbf{ASIA} & asia\_osm & 11.95 & 25.42 & $\SI{1.78e-05}{}$ & $\SI{0.33}{\giga\byte}$ \\
		\textbf{RC} & road\_central & 14.08 & 33.87 & $\SI{1.71e-05}{}$ & $\SI{0.43}{\giga\byte}$ \\
		\textbf{WK} & Wikipedia & 3.56 & 45.00 & $\SI{3.55e-4}{}$ &  $\SI{0.60}{\giga\byte}$ \\
		\textbf{HT} & hugetrace-00020 & 16.00 & 47.80 & $\SI{1.87e-05}{}$ & $\SI{0.61}{\giga\byte}$ \\
		\textbf{WB} & wb-edu & 9.84 & 57.15 & $\SI{5.90e-05}{}$ & $\SI{0.73}{\giga\byte}$ \\
		


		\bottomrule
	\end{tabular}
    }
    \label{tab:matrices}
\end{table}

To prove that our custom FPGA design is suitable for solving large-scale Top-K sparse eigenproblems, we compare it against the popular ARPACK library, measuring how it compares in terms of execution time, power efficiency, and accuracy.
The multi-threaded ARPACK library \cite{lehoucq1998arpack}, a Top-K sparse eigensolver that employs \gls{iram}, runs on two Intel Xeon Gold 6248 (80 threads in total) and 384 GB of DRAM using single-precision floating-point arithmetic.
Our eigensolver is prototyped on a Xilinx Alveo U280 accelerator card equipped with 8 GB of
HBM2 memory, 32GB of DDR4 memory, and a \texttt{xcu280-fsvh2892-2L-e} FPGA whose resources are reported in \Cref{tab:resources}.
Results are averaged over 20 runs.

Tests are carried out using a collection of large sparse matrices representing graph topologies, each containing millions of rows and non-zero entries (\Cref{tab:matrices}).
All test matrices come from the SuiteSparse collection \cite{davis2011university}.
While our evaluation is focused on sparse matrices representing graphs, our Top-K sparse eigenproblem FPGA design is applicable to other domains such as image analysis \cite{perona1998factorization,shi2000normalized,tung2010enabling}.

Resource utilization and clock frequency of our design are reported in \Cref{tab:resources}. 
The Lanczos algorithm and Jacobi algorithm have similar utilization, with around 20\% LUT utilization each (50\% of the available LUTs in each \gls{slr}). Although the \gls{sa} architecture of the Jacobi algorithm processes small $K \times K$ inputs, it requires the computation of many trigonometric operations and multiplications (16 and $>$ 800 for $K\!=\!8$) in each iteration. Resource utilization of the Jacobi algorithm scales quadratically with the number of eigenvalues $K$, while the Lanczos algorithm is not affected.

\begin{figure*}[t]
    \centering
    \includegraphics[width=\linewidth]{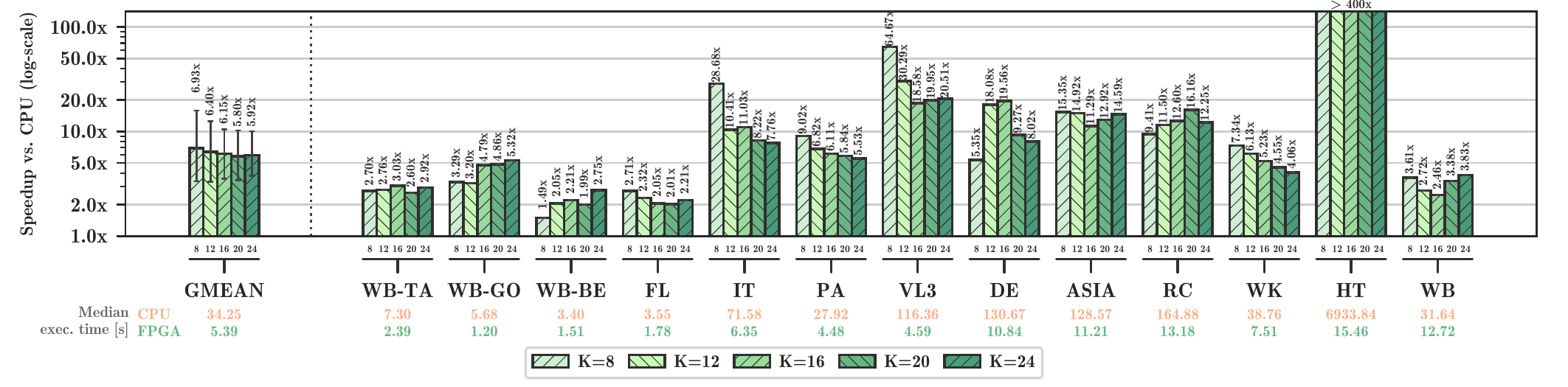}\\
    \caption{Speedup (higher is better) of our Top-K sparse eigensolver vs. the ARPACK multi-core CPU library, 
    for increasing number of eigenvalues $K$. Geomean excludes the outlier graph HT, where the speedup of our FPGA design exceeds 400x.}
    \label{fig:exec_time}
\end{figure*}


\begin{figure}%
    \centering
     \subfloat[\centering]{
    {\includegraphics[width=0.45\columnwidth]{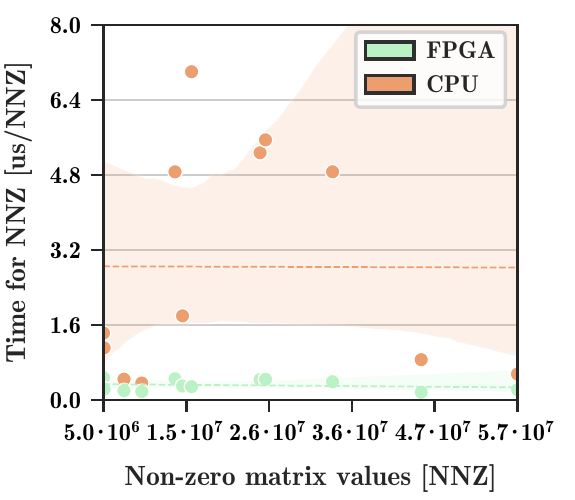}
    \label{fig:correlation}}
    }
    \subfloat[\centering]{
    {\includegraphics[width=0.45\columnwidth]{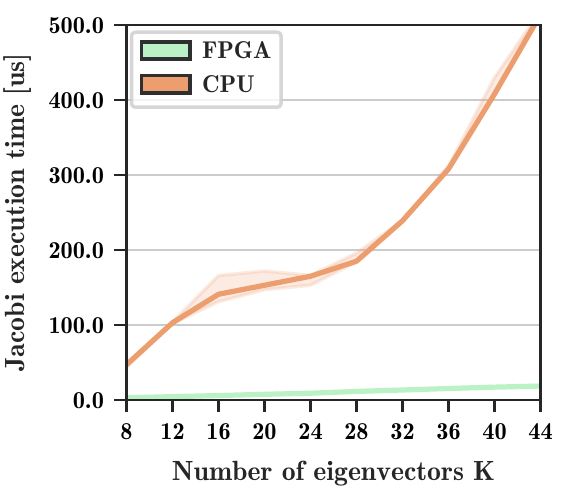}
    \label{fig:jacobi_exec_time}}
    }
    \caption{\Circled{a} Relation between number of matrix non-zero values and time to process a single value. \Circled{b} Speedup vs. CPU of our Systolic Array architecture for the Jacobi algorithm.}%
    \label{fig:jacobi_and_correlation}%
\end{figure}

\begin{figure}[t]
    \centering
    \includegraphics[width=\columnwidth]{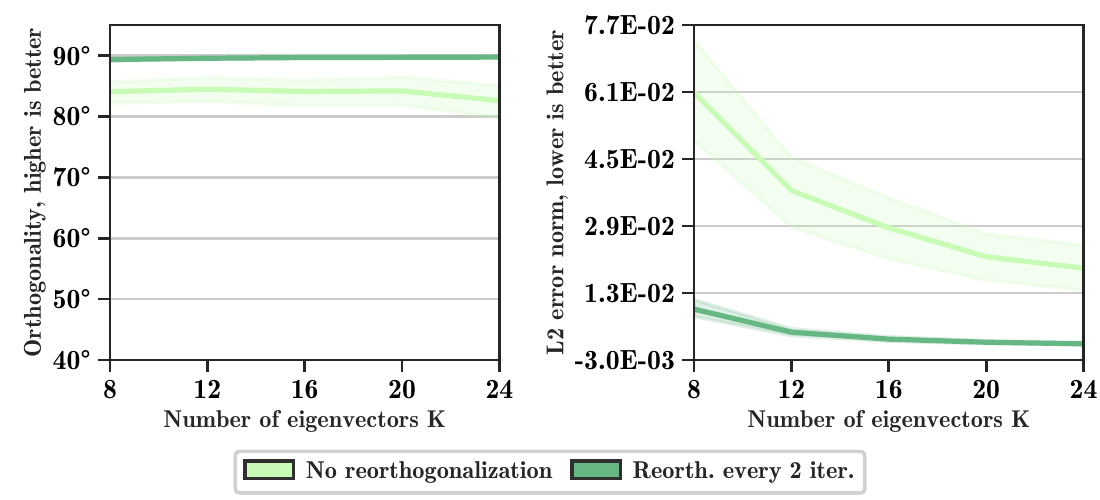}\\
    \caption{Accuracy of our Top-K sparse eigensolver, in terms of orthogonality and reconstruction error, for increasing $K$.}
    \label{fig:accuracy}
\end{figure}

\subsection{Execution Time}\label{sec:exec_time}

We measure the execution time speedup of the FPGA-based hardware design implementing our Top-K sparse eigenproblem solver against the CPU baseline and report results in \Cref{fig:exec_time}.
We are always faster than the baseline, with a geometric mean speedup of 6.22x, up to 64x for specific graphs. 
The speedup is mostly unaffected by $K$, showing how our design can efficiently compute many eigenvalues at once. 
\Cref{fig:jacobi_and_correlation}A shows how the time required by our FPGA design to process a single matrix value is unaffected by the overall graph size, while the CPU behavior is drastically more unpredictable.

We estimate that the Lanczos dominates the overall execution time due to the \gls{spmv} computations, taking more than 99\% of the execution time.
However, optimizing the Jacobi algorithm with a \gls{sa} design is still worth the effort, compared to running this step on CPU.
\Cref{fig:jacobi_and_correlation}B shows the speedup of our Jacobi \gls{sa} design compared to an optimized C++ CPU implementation: the execution time on CPU grows quadratically due to repeated matrix multiplications, becoming a non-negligible part of the execution time for large $K$. 

Our hardware design synthesized at 225 Mhz on the Alveo U280 accelerator card. A clock frequency beyond 225 Mhz does not significantly improve performance as \gls{spmv} represents the main computational bottleneck in the computation, and its performance is bound by HBM bandwidth \cite{lu2021demystifying}. Each \gls{spmv} \gls{cu} processes data at the maximum bandwidth offered by the HBM channel from which it reads the matrix (14.37 GB/s, for a total of 71.87 GB/s using 5 \gls{cu}).

\subsection{Power Efficiency}\label{sec:power_efficiency}

We measured via an external power meter that our FPGA design consumes about 38W during execution, plus 40W for the host server. 
The CPU implementation consumes around 300W during execution. Our FPGA design provides 49x higher Performance/Watt ratio (24x if accounting for the FPGA host machine): we provide higher performance without sacrificing power efficiency, making our design suitable for repeated computations typical of data center applications.

\subsection{Accuracy Analysis of the Approximate Eigencomputation}\label{sec:accuracy}
The Lanczos algorithm is known to suffer from numerical instability \cite{paige1972computational}. To limit this phenomenon, we reorganize the algorithm's operations as in \cite{paige1972computational} and apply reorthogonalization as in \cite{parlett98}.
To assess the stability of our design, we measure the eigenvectors' pairwise orthogonality and the eigenvector error norm. 
Eigenvectors must form an orthonormal basis and be pairwise orthogonal, i.e. their dot product is 0, or equivalently their angle is $\pi / 2$. For each pair of eigenvectors, we measure the average angle that occurs.
Then, if $\lambda$ is an eigenvalue of $\mathbf{M}$ and $v$ is its associated eigenvector, it must hold $\mathbf{M}v = \lambda v$.
By measuring the L2 norm of $\mathbf{M}v - \lambda v$ for all $v$ we evaluate how precise the eigenvector computation is.
Results are reported, for increasing $K$, in \Cref{fig:accuracy}, aggregated on all graphs due to space constraints.
Accuracy is excellent if reorthogonalization is applied 
every two iterations, but even without this procedure results are satisfactory.
Despite using fixed-precision arithmetic in the Lanczos algorithm, the average reconstruction error is below $10^{-3}$, and the average orthogonality is $>89.9$ degrees, when applying reorthogonalization every two iterations.
Orthogonality is not affected by $K$, while the average reconstruction error improves as $K$ increases.
Spectral methods in machine learning applications use eigenvectors to capture the most important features of their input and do not usually require the same degree of precision as other engineering applications.
Reorthogonalization adds an overhead up to $\mathcal{O}(nK^2/2)$ to the algorithm compared to \Cref{fig:exec_time}. On large graphs this overhead is negligible compared to \gls{spmv}, and is a viable option in applications where maximum accuracy is necessary.
Still, our hardware design can provide excellent accuracy while being significantly faster than a highly optimized CPU implementation.

\section{Conclusion}\label{sec:conclusion}

The computation of the Top-K eigenvalues and eigenvectors on large graphs represented as sparse matrices is critical in spectral methods, a class of powerful Machine Learning algorithms that can extract useful features from graphs.
We solve the Top-K sparse eigenproblem with a new algorithm that is optimized for reconfigurable hardware designs: in the first part of the computation, we exploit the enormous bandwidth of HBM through the Lanczos algorithm, while in the second part, we introduce a systolic array architecture that efficiently parallelizes the compute-intensive Jacobi eigenvalue algorithm.
Compared to the popular ARPACK CPU library, we achieve a geomean speedup of 6.22x on 13 graphs with millions of vertices, raising the bar for high-performance Top-K sparse eigensolvers at a large scale.

As future work, we will extend our hardware design to support non-Hermitian matrices through the Implicitly Restarted Arnoldi Method.
We will also investigate heterogeneous implementations that combine the abundant memory bandwidth of GPUs for high-performance \gls{spmv} with our systolic array FPGA design for the Jacobi eigenvalue. 
\bibliographystyle{IEEEtran}
\bibliography{bibfile.bib}

\end{document}